# Identifying regions of interest in whole slide images of renal cell carcinoma

Mohammed Lamine Benomar[1,2] · Nesma Settouti[1] · Eric Debreuve[3] · Xavier Descombes[3] · Damien Ambrosetti[4]

**Abstract**

**Purpose** The histopathological images contain a huge amount of information, which can make diagnosis an extremely time-consuming and tedious task. In this study, we developed a completely automated system to detect regions of interest (ROIs) in whole slide images (WSI) of renal cell carcinoma (RCC), to reduce time analysis and assist pathologists in making more accurate decisions.

**Methods** For this purpose, the WSIs are divided into patches at high resolution and a method is proposed to classify the patches into a tumor and healthy tissue. The proposed approach is based on an efficient texture descriptor named dominant rotated local binary pattern (DRLBP) and color transformation (hematoxylin and violet channels) to reveal and exploit the immense texture variability at the microscopic high magnifications level. Thereby, the DRLBPs retain the structural information and utilize the magnitude values in a local neighborhood for more discriminative power. For the classification of the relevant ROIs, feature extraction of WSIs patches was performed on the color channels separately to form the histograms. Next, we used the most frequently occurring patterns as a feature selection step to discard non-informative features. The performances of different classifiers (k-NN, SVM and RF) on a set of 1800 kidney cancer patches originating from 12 whole slide images were compared and evaluated. Furthermore, the small size of the image dataset allows to investigate deep learning approach based on transfer learning for image patches classification by using deep features (VGG-16) and fine-tuning (ResNet-50) methods.

**Results** High recognition accuracy was obtained and the classifiers are efficient, the best precision result was 99.17% achieved with SVM. Moreover, transfer learning models perform well with comparable performance, and the highest precision using ResNet-50 reached 98.50%. The proposed approach results revealed a very efficient image classification and demonstrated efficacy in identifying ROIs.

**Conclusion** This study presents an automatic system to detect regions of interest relevant to the diagnosis of kidney cancer in whole slide histopathology images.

## Introduction

The renal cell carcinoma (RCC) represents about 90% of the kidney cancers according to the World Health Organization (WHO), with several hundreds of thousands of new cases each year (Jonasch et al. 2014). In 2016, the WHO has updated the categorization of subtypes of renal cell tumors (Moch et al. 2016). Only two of them are benign (papillary adenoma and oncocytoma). Among the malignant tumor types, the most common ones are the clear cell renal cell carcinoma (ccRCC, 75% of the cases), the papillary carcinoma (pRCC, 10%), and the chromophobe carcinoma (5%) (Muglia and Prando 2015).

Conventional light microscopy has been a primary tool for RCC diagnosis and prognosis evaluation for decades by pathologists using histopathologic information derived from the identified tumor in the partial or total kidney nephrectomy. This analysis is based on the cellular morphology and

✉ Mohammed Lamine Benomar
mohamedamine.benomar@univ-tlemcen.dz

Extended author information available on the last page of the article.

the architecture of the tumor (primarily related to the vascular organization of the tissue), thus implying an exhaustive microscopic exam of the entire slide (Sabo et al. 2001; Cheville et al. 2003). It should be noted that a pathologist diagnosis on a histology slide requires several hours, whether to confirm the tumor nature of a suspect tissue area previously identified by other imaging methods, to characterize a tumor and define histological subtypes to optimize the associated therapeutic management, or to qualify the tumor margins associated with surgical resection.

Nowadays, the automatic whole slide imaging (WSI) analysis development is rapidly gaining popularity among physicians by introducing several novel tools and applications for digital pathology slides, including computer-aided diagnosis and virtual microscopy (Velez et al. 2008; Weinstein et al. 2009; Wilbur et al. 2009; Fallon et al. 2010). The WSI brings new challenges to image processing and offers a digital copy of an entire histological slide instead of conventional microscopy scanning that allows only a view of a fraction of a slide at a time. This approach, called whole slide pattern recognition, contributes significantly to improve clinical diagnosis and provides a more accurate quantitative evaluation of cellular material than traditional pathology slide analysis (Yeh et al. 2014b).

However, the analysis of histological images poses several difficulties; as we can see in Fig. 1, the histological images are compressed and put in whole slide high-resolution imaging format (Ho et al. 2006; Melo et al. 2020). WSI is mostly background and contains non-tumor (debris, adipose tissue (fat), mucus, normal mucosa, stroma, etc.) and tumor tissue. For subtype classification, only the tumor areas are interesting, it is necessary to segment these regions of interest (ROI) beforehand, to eliminate the other areas of the sample (healthy tissue) which have no interest in the analysis of the tumor and would not provide any relevant information to the pathologist. This segmentation can be performed at any level of magnification of WSI images, but due to the large size of the images at maximum magnification, a reasonable compromise between computation time and quality of segmentation is to be found.

Segmentation of renal tissue is a field that has been lightly explored in histological image processing. Most works can be divided on two major axes: *tumor region segmentation* (Cheng et al. 2017; de Bel et al. 2018; Srinidhi et al. 2019; Hossain and Sakib 2020) and *tumor grading* (Yeh et al. 2014a; Tian et al. 2019; Delahunt et al. 2019; Delahunt et al. 2013). Among the proposed approaches, most are oriented towards supervised learning (Apou et al. 2014; Wang 2011). This requires the prior intervention of an expert to manually annotate certain pixels (or certain regions) of the image to constitute a learning set on which the classification algorithm can learn.

It should be noted that the evaluation and annotation of stained slides tissue remain tedious, time-consuming, and prone to error for pathologists. Strength of this finding, Fuchs et al. (2008) proposed a novel weakly supervised classification method, which is based on an iterative morphological filtering algorithm and a soft margin SVM, a semi-supervised classification for cell detection and segmentation.

In Yeh et al. (2014a), a nuclei segmentation algorithm is proposed based on the analysis of the spatial distribution of nuclear size, which retains spatial information that can be leveraged to facilitate locating regions of interest. A support vector machine is applied to nuclei recognition using an interactive interface to manually select regions (e.g., nuclei, cell body, background tissue) on 39 hematoxylins and eosin-stained digitized slides of clear cell RCC with varying grades. The sizes of the recognized nuclei were estimated, and kernel regression was used to estimate the spatial distribution of nuclear size across the entire slides.

Other approaches based on unsupervised learning (or clustering) were proposed to automate the detection of areas of interest (ROI). Zubiolo (2015) starts from the observation that tumor areas are darker and more heterogeneous (due to the presence of nuclei, appearing in blue-black thanks to hematoxylin staining). This heterogeneity is quantified by three measures: the local entropy of the image, the variance, and the median in the vicinity of a pixel. Subsequently, the unsupervised $K$-means classification algorithm is used to segment tumor areas for different values of $K$ ($k = 3$ and $k = 4$), with different initialization followed by a majority voting step for the choice of the partition of a cluster. Finally, the class chosen to represent the region of interest (ROI) is the one that maximizes the mean entropy. The segmentation results show some limitations of the proposed method due to the complexity and inter-image variability (different morphology and coloration), as well as intra-image variability (variable network density, different tissue components, etc.).

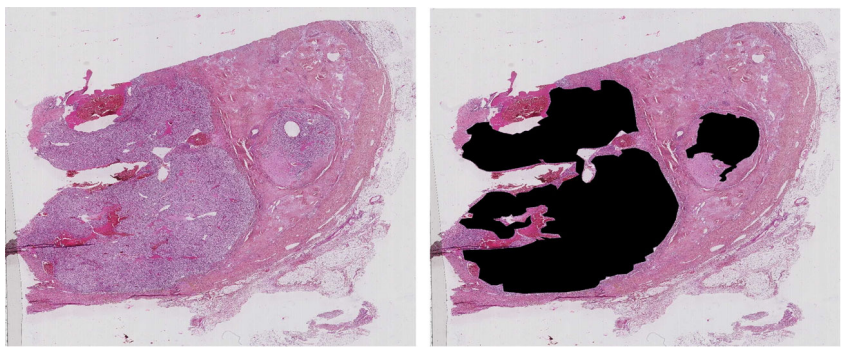

**Fig. 1** Histopathology whole slide image: (left) original WSI, (right) ground truth image with the ROIs indicated in black (pathologists' annotation)

Cheng et al. (2017) aim to improve the prognostic prediction of papillary RCC through objective features derived from a cohort of 190 histopathology patients images with papillary renal cell carcinoma obtained from The Cancer Genome Atlas project. A fully automated method is used to learn potential nucleus patterns via an unsupervised feature learning algorithm followed by clustering. The proposed workflow consists of two modules. The first module is learning nucleus patterns using stacked sparse auto-encoder. The second one, generating topological features (bag of edge histogram features BOEH) of an image using the learned nucleus patterns and Delaunay triangulation.

In recent years, the convolutional neuronal networks (Fukushima 1980; Waibel et al. 1989; Lecun et al. 1998) achieved significant performance on computational histopathology (Srinidhi et al. 2019; Dimitriou et al. 2019). Tabibu et al. (2019) demonstrate that a deep learning framework can be a good candidate for automatic classification of clear cell, chromophobe and papillary renal cell carcinoma (RCC) subtypes on The Cancer Genome Atlas (TCGA) slides that contains over 11000 cases. The lack of enough normal tissue samples and the class imbalance caused were the major challenges, where a two-step procedure was followed by data augmentation: random vertical flip, rotation (25 to + 25°), noise addition, and then a weighted re-sampling technique. In Hossain and Sakib (2020), synthetic but annotated renal cell nuclei data are generated based on non-synthetic data reference, to tackle the need for a large amount of annotated data required for a deep learning network. The proposed approach generates synthetic nuclei patches close to non-synthetic reference patches and measures the performance of the U-net segmentation network. The initial synthetic patches are refined with a SURF feature learning algorithm to score each synthetic patch. Convolutional neural networks (CNNs) were able to identify the tumorous tissue patterns as well as the inherent texture differences among RCC subtypes. Further, morphological features were extracted from high probability tumor regions identified by the CNN to predict patient survival outcome of the most common clear cell RCC. However, the classification of the various stages and grades of tumors did not lead to satisfactory results. de Bel et al. (2018) proposed a two-step approach to perform a structure segmentation and subsequently an immune cell detection to quantify tubular inflammation. Automatic multi-class instance segmentation is holding out, with a total of seven structures of the kidney anatomy for the segmentation task. A modified U-net was developed with the addition of a second decoder before the fourth max-pooling layer for structure and border segmentation. This case study was severely limited by the small dataset of five WSIs where some classes had few annotations leading to a lower Dice score.

Lu et al. (2020) demonstrated the feasibility and effectiveness of federated learning that offers ways to attenuate the immense heterogeneity of histological data (different patient groups corresponding to histology specimens, variations in tissue preparation, various fixation approaches and staining protocols, different scanners model used for scanning, etc.) combined with weakly supervised multi-instance learning to histological sub-typing of breast and kidney cancer classification using only slide-level labels for supervision.

Through the different application of convolutional neuronal networks in histopathology, we can note that it faces several challenges: First, deep CNNs require a large amount of annotated data to achieve a good performance, which is a limiting factor in histopathology (there is a lot of data but not all of it is labeled). Second, when deep networks are trained with few data, they are prone to "over-fitting," as they cannot generalize very well to test data. Third, deep CNNs require massive computational resources to learn the model which usually requires a long dedication of many professionals.

To overcome the problem of CNNs' dependence on annotated data, many recent works have focused on transfer learning for automated classification of histopathology images (Xu et al. 2019; Talo 2019; Patil et al. 2020). Indeed, Transfer Learning is based on the simple idea of reusing a deep learning model already learned on a large database for a problem of smaller dimensions. It also has the advantage of being inexpensive compared to training a complete deep neural network on a large database. As presented by Li and Plataniotis (2020) work, they experimentally investigated and reported transfer efficiency of deep net's form nature representation over different pathology image sets.

In this context, we can distinguish several approaches depending on what we want to transfer, when and how we want to transfer it. Overall, we can distinguish 2 types of strategies: the first one is deep features extraction as experimented by Yousefi and Nie (2019) and Alinsaif and Lang (2020). The idea is to reuse a pre-trained network without its final layer. This new network then works as a fixed feature extractor for other tasks. The second strategy is fine-tuning, in which not only the last layer is replaced to achieve classification, but other layers are also selectively re-trained. The idea is therefore to freeze (that is to say, fix the weights) of certain layers during the training and to refine the rest to answer the problem. Indeed, deep neural networks are highly configurable architectures with various hyper-parameters. Additionally, while the first layers capture generic characteristics, the later layers focus more on the specific task at hand. Bayramoglu and Heikkilä (2016) initiated a comparative study to investigate the question of it is interesting to use transfer learning and fine-tuning in biomedical image analysis and especially cell nuclei classification in histopathology images to reduce the effort of manual data labeling and still obtain a full deep representation for the target task? They compared four different CNN models with depths ranging from 3 to 13 convolutional layers. Empirical results show that initializing network parameters with transferred features can improve classification performance for any model. However, deeper architectures trained on larger datasets converge quickly.

In this paper, we present an automatic WSI analysis method to localize the tumor regions across an entire histological slide by considering the important properties of the histopathology images namely texture and colorimetry features on a set of Kidney whole slides images. This work aims to quickly locate high-grade regions of interest (ROIs) to determine the tumor subtype and to facilitate grading of the renal cell carcinoma (RCC) for pathologists (Fuhrman et al. 1982; Ficarra et al. 2005; Sun et al. 2009; Hong et al. 2011; Yeh et al. 2014a). We proposed a simple and robust approach based on DRLBP (dominant rotated local binary pattern) (Mehta and Egiazarian 2016) features extraction from different color channels (H and V) to brings pertinent texture information and a features selection step to provide a final feature vector that can well describe the texture characteristics of histological images. To build a decision function for the binary patches classification as tumor or not-tumor, we opt for current approaches of the literature in the context of learning a single classifier as generative method k-NN and discriminative method SVM, we applied also the homogeneous ensemble method based on decision trees Random Forest. We pushed the reflection by studying the contribution of Deep learning methods through transfer learning with Fine-tuning and Deep Feature extraction strategies based on pre-trained models (ResNet-50 (He et al. 2016) and VGG-16 (Simonyan and Zisserman 2015)) from the ImageNet dataset (Deng et al. 2009) to identify regions of interest in WSI of renal cell carcinoma.

The paper is organized as follows: renal histopathological slide analysis methods are presented in the "Methods" section. The results are presented in the "Results" section followed by a discussion in "Discussion" section. Finally, conclusions from this study and possible future works are presented in the "Conclusion" section.

## Methods

This study proposes a segmentation of the tumor regions in histopathological whole slide images in three main steps: preprocessing, features extraction, and classification using machine learning algorithms. Our proposed identification ROIs scheme is illustrated with an example of classifying image patches of the histological WSI as tumor or not-tumor (binary classification) based on sliding window approach as shown in Fig. 2, where the images were treated patch-wise as well as slide-wise. The goal is to train a model that best predicts the label for an input test image patches based on image-level labels annotated by the pathologist (tumor / not-tumor).

Firstly, in the preprocessing step, the entire whole slide images were tessellated into local mini patches of $600 \times 600$ pixels to analyze each patch independently for ROI detection. In the features extraction step, color transformation and texture information have been used for machine learning tasks. Two techniques were employed, which are color transformation (a.k.a. channel combination H and V) making the tumor nucleus regions and vascular networks more clearly distinguishable in the patches, followed by DRLBP pattern extraction since texture features play an important role in histopathological image analysis and considering the discriminative power of DRLBP. Finally, patches are classified as tumor or not-tumor to identify high probability tumorous regions, the result consists of a binary mask for the tumor regions in the WSI where small isolated patches are removed using morphological operations.

### Preprocessing

Histological slides are mostly stained using eosin and hematoxylin dyes (Layton et al. 2019), since the eosin stains pink the cytoplasm and the hematoxylin is a base that colors the nucleus blue/violet. Figure 3 shows how many details are contained in a very small portion of the image.

The computational and memory difficulties encountered in the WSI approach, because of its huge dimensions where images have up to a billion pixels and files size tends to be larger than 1 GB, requires at first, a division of the whole slide image into equal-sized sub-images (in our case: $600 \times 600$ patch size) on which local information can be extracted with more basic operations, such as thresholding, filtering and morphological methods (Zubiolo et al. 2016). For all histological slides, we followed the framework depicted in Fig. 2. Then, in a second step, we combine the individual patch classification results to determine the ROIs across an entire pathology slide. This strategy allows the detection of ROIs to be carried out at a reduced given scale.

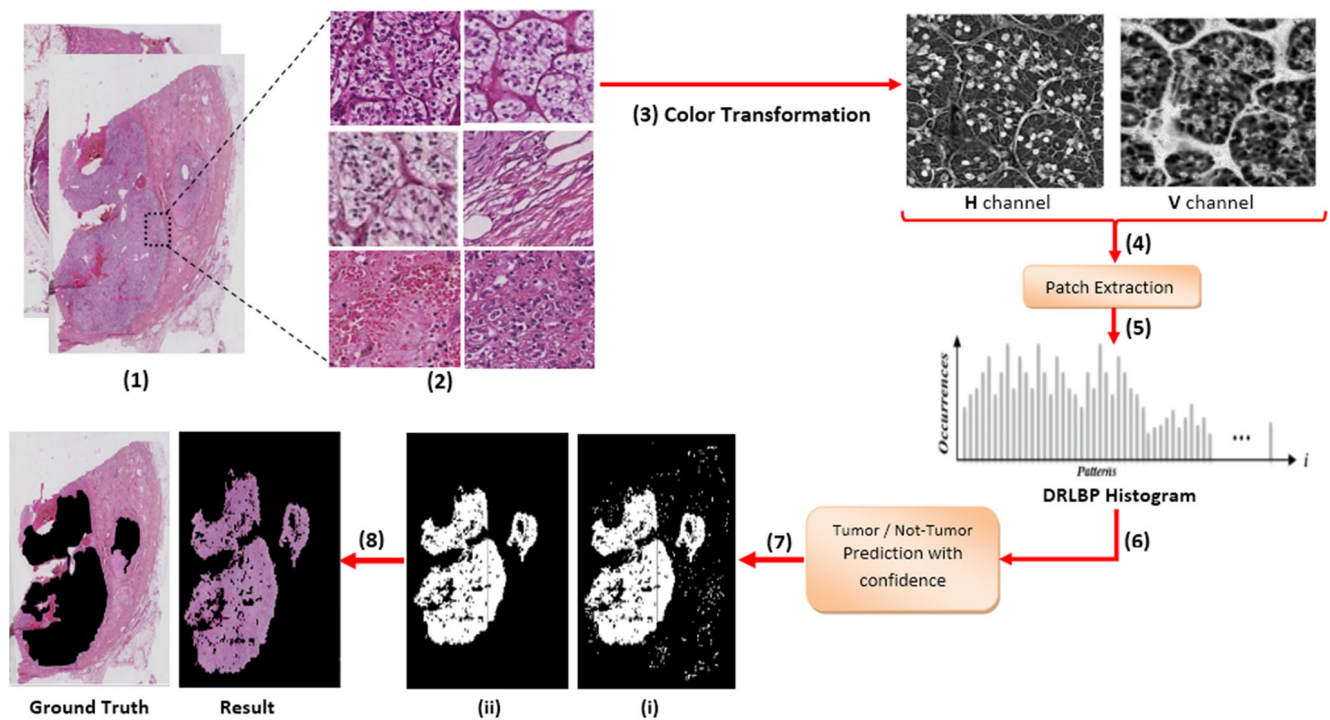

**Fig. 2** Tumor segmentation pipeline. (1) Kidney WSIs (whole slide images). (2) WSIs are split into 600 × 600-pixel patches which constitute a dataset. (3) Color transformation for a vascular network (V channel) and nucleus pixels enhancement (H channel). (4) Patch extraction based on nuclei ratio. (5) DRLBP features extraction and selection from H and V channels, then histogram concatenation (6) Binary classification (patches classified as tumor or not-tumor). (7) High-probability patches identified by the trained algorithm, (i) binary mask generated where each pixel of the map corresponds to a patch (600 × 600 pixels) in the input WSI, (ii) Small patches removed using morphological operations. (8) The resulting image is then displayed by a sliding window to obtain the final tumor map for the entire whole slide image

## Color transformation

A recurrent difficulty with microscopic histopathology images is color variations due to the variety of the materials (scanner models), the several dye manufacturers, and the staining procedure. Different markers are used to highlight objects in a histological image; therefore a markers separation step is required to facilitate detection and segmentation. In such cases, color transformation is mainly used, where the original image must be transformed into a new color space according to the marker intensity (Alsheh Ali 2015).

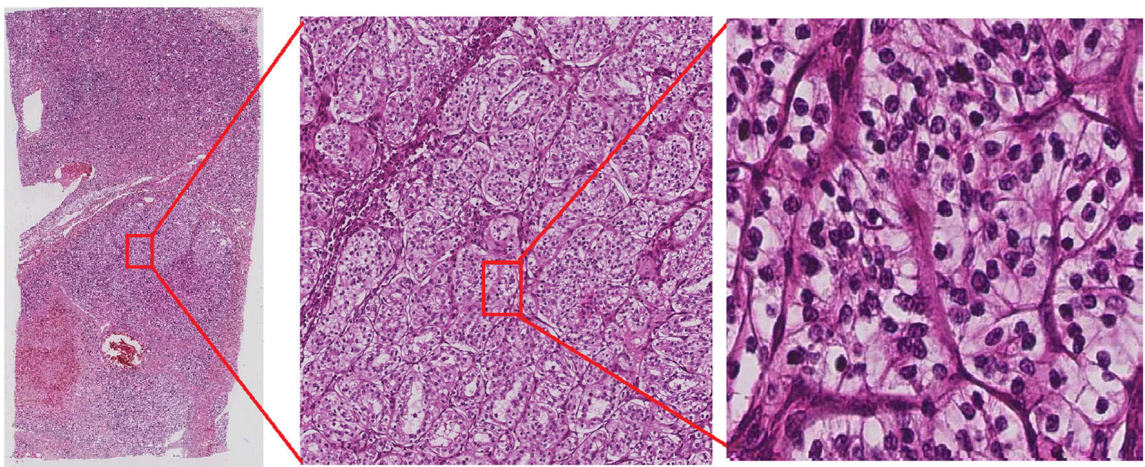

**Fig. 3** Zoomed-in on the red box and the differents magnification level: leftmost image shows the whole slide image. rightmost image presents highest magnification level and clearly shows the nuclei region (the violet network is vessels and the blue/violet dots are cell nuclei)

Cell shape and nucleus morphology analysis helps in the diagnosis of kidney tumors. Furthermore, the vascular network grows during tumor formation. Consequently, the topology and geometry of the vascular network is also a crucial diagnostic criterion for the tumor histological type and grade (Zubiolo et al. 2016). The goal is to make the vascular network and the tumor nucleus more clearly visible to segment tumor regions in the whole slide image (WSI). Since the vessels appear in purple in the RGB images (Fig. 4), we used the average of the red and the blue channels, known as the V channel (V for violet) on which the vascular structures clearly appear:

$$V = 0.5 \frac{R + B}{\sqrt{R^2 + G^2 + B^2}} \qquad (1)$$

In addition, the H channel, corresponding to hematoxylin was applied to highlight the nuclei:

$$H = \frac{R}{C_3} \; where \; C_3 = arctan\left(\frac{B}{max(R, G)}\right) \qquad (2)$$

It is important to note that the H and V channels are not to be confused with the hue and saturation of the HSV color space, and although the H channel does correspond to hematoxylin, the V channel is distinct from eosin, despite what one might be tempted to conclude. In summary, the color transformation used in our approach are:

– the V channel, corresponding to the violet channel (Eq. 1) highlight the vascular network,
– the H channel, corresponding to hematoxylin (Eq. 2) emphasis the nuclei.

### Patch extraction

Due to the huge amount of information contained in the histological image, the individual patches generated from slides may not contain relevant information since they may be primarily backgrounded (white areas, debris, adipose tissue (fat), blood, etc.) patches and lack any significant tumor patterns (see the leftmost image in Fig. 3 and the last line images in Fig. 5). Nevertheless, for the clinical diagnosis and prognostics of carcinomas, various parameters are calculated, among them, the overall cellularity (nuclei ratio) wish is a critical and challenging task. Indeed, uncontrolled cell nuclei growth, mainly observed in areas where the presence of cell nuclei is abnormally high, is a common sign of carcinomas. Hence, the nuclei ratio (cellularity) can be used to filter out most of the healthy tissue patches according to their cell nuclei amount by assigning high cellularity to high-grade carcinoma and select patches with a higher probability of cancer in the whole slide image (Travis 2014; Riasatian et al. 2021) to avoid biasing the classifier and locate the tumor regions within a reasonable processing time. Therefore, as a preprocessing step, a nuclei segmentation function was implanted to measure the cell nuclei ratio of each patch; first, the color transformation is applied to convert the initial RGB color space to hematoxylin (H channel) using the formula (2) where the nucleus appears more clearly (see Fig. 5); then, morphological opening (using disk-shaped structuring element), empirical thresholding, and morphological closing are successively applied to get the binary nucleus mask of each patch (Zubiolo et al. 2016). Finally, the cell nuclei ratio of each patch is obtained by averaging the nucleus segmentation areas over the patch area (600 × 600 pixels), then patches are ranked according to their ratio, patches with less than 3% are rejected as learning cases.

### Features extraction

In a computer-aided system (CAD), the feature extraction step is generally considered the most important step. One of the important low-level characteristics of histopathology images is the texture, which can be considered as a similarity criterion for image classification where the specificity and repeatability of texture patterns can significantly increase the system performance (Erfankhah et al. 2019).

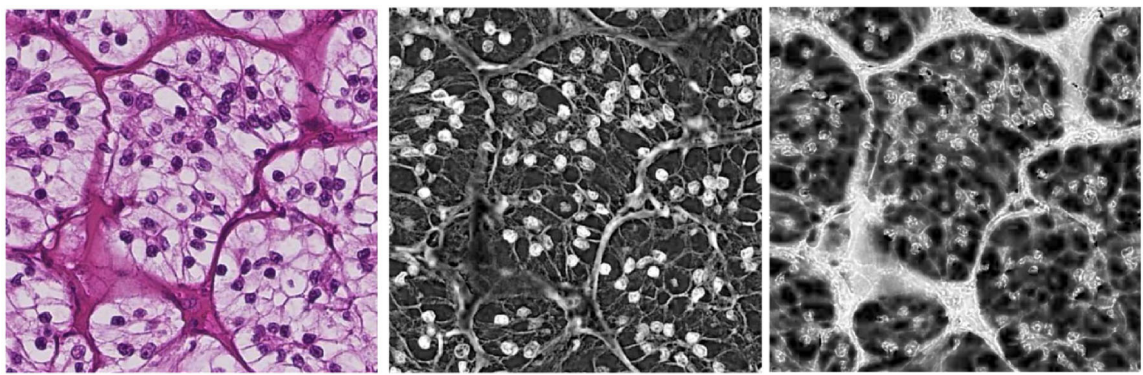

**Fig. 4** Color transformation. (from left to right) Input RGB patch, H channel (used for nucleus extraction), V channel (where the vascular network has the brightest aspect)

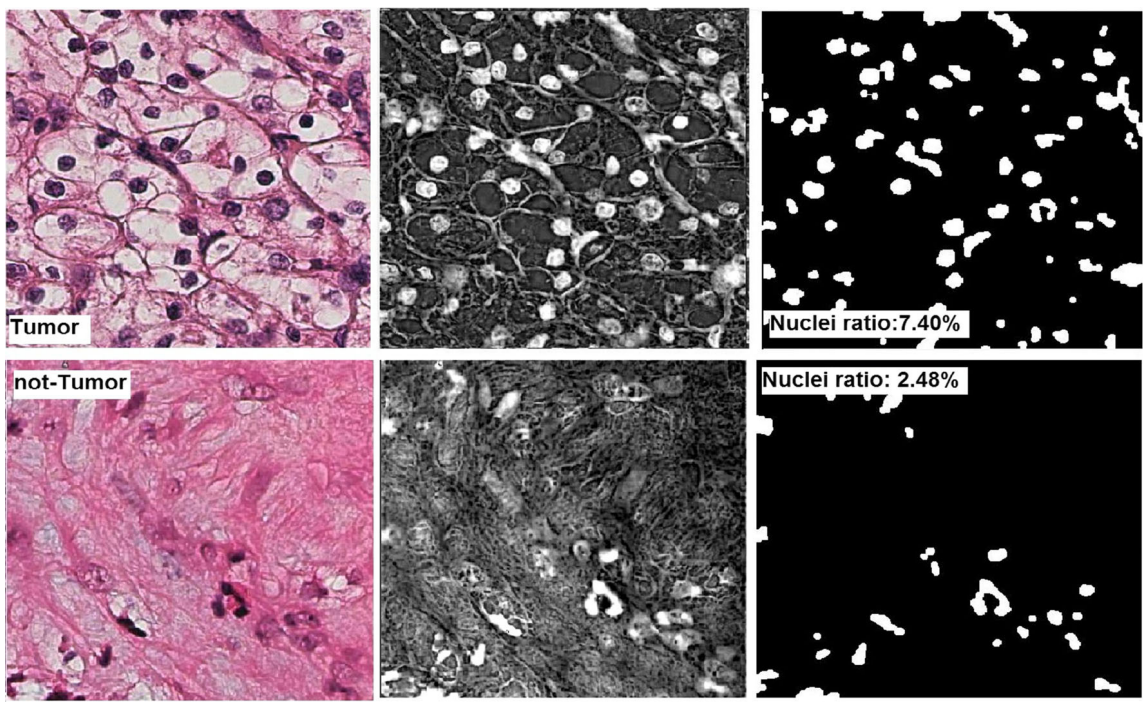

**Fig. 5** Examples of cell nuclei segmentation. (From left to right) RGB patches, H channel, binary mask result

In the literature, there are different efficient methods for feature extraction, including the local binary patterns (LBP) (Ojala et al. 2002), the scale-invariant feature transform (SIFT) (Lowe 2004), speeded up robust features (SURF) (Bay et al. 2008), and histograms of oriented gradients (HoG). Nevertheless, LBP may be better qualified as a texture operator due to computational simplicity and high discriminative performance (Mehta and Egiazarian 2016). Therefore, LBP performs well when used for histopathological image classification and retrieval (Öztürk and Bayram 2018) by sampling all of the pixels (a.k.a. dense sampling) in the image and creating the LBP histogram. However, the remaining operators (SIFT, SURF, and HoG) are efficient especially in object detection, face recognition, and tracking applications (Erfankhah et al. 2019).

In this paper, we applied an extension of LBP called DRLBP (dominant rotated local binary pattern) (Mehta and Egiazarian 2016) which incorporates the complete structural information and the rotation invariance property to overcome the limitations of the traditional LBP (Ojala et al. 2002). Indeed, rotation invariance is generally required in medical image classification since the element appears at various angles depending on the camera rotation or the self-rotation of the captured objects.

The LBP descriptor computation to extract binary patterns in a local circular region is based on the difference between the central pixel and the surrounding neighbors then combining the signs of these differences using fixed-order unique weights to compute the final descriptor (see Fig. 6). Hence, the LBP operator provides very different values for a simple image rotation, and the information regarding the magnitude of differences is completely neglected. The equation can be written as:

$$LBP_{R,P} = \sum_{p=0}^{P-1} s(g_p - g_c).2^p, \quad s(g_p - g_c) = \begin{cases} 1 : g_p \geq g_c \\ 0 : g_p < g_c \end{cases} \quad (3)$$

where the pixels $p$ are defined by a circular neighborhood set of radius $R$ ($R \in \mathbb{N}$) and cardinality $P$ (see Fig. 6), $p$ is the sampling points. $g_c$ and $g_p$ corresponds to intensity value of central pixel and its neighbor respectively (Ojala et al. 2002).

To obtain rotation invariance, the dominant direction ($D$) is defined as the index of the neighboring pixel whose difference from the central pixel is maximum (Mehta and Egiazarian 2016):

$$D = \arg \max_{p \in 0, 1 \dots P-1} |g_p - g_c| \quad (4)$$

The RLBP (rotated local binary pattern) descriptor is computed by rotating the weights with respect to $D$. It is defined as follows:

$$RLBP_{R,P} = \sum_{p=0}^{P-1} s(g_p - g_c).2^{mod(p-D,P)} \quad (5)$$

**Fig. 6** (a) The surrounding neighbors for LBP. (b) Associated weights

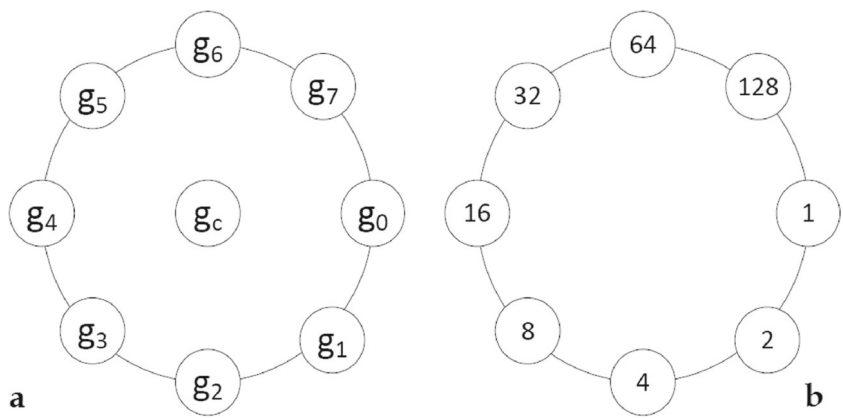

where modulus (mod) operator circularly shifts the weights for $D$ and the weights sequence is maintained (Mehta and Egiazarian 2016). Therefore, by using this method we also incorporate supplementary information from the local neighborhood to increase the discriminative power of the operator without increasing dimensionality since a subset of patterns is selected according to their distribution in the training images (Mehta and Egiazarian 2016). Increasing the value of $P$ (number of neighboring pixels) improves the performance of the descriptor but simultaneously increases the dimension of the features.

Thereby, after the binary patterns extraction from each color channel space separately (i.e., $H$ channel and $V$ channel) and the construction of the histograms to form the feature representation of each patch. Moreover, to highlight the different patterns of the image, Fig. 7 shows the histogram distributions of the tumor against not-tumor patch patterns from the V (violet) channel, the histograms are different for tumor and not-tumor patches. Hence, tumor patches reveal peaks representing groups of continuous pixels that can be interpreted as edges, in contrast to the not-tumor patch, the histogram show peaks at largest values considered as flat regions (when surrounding pixels are all black or all white). Next, the second step is a simple but powerful method of feature selection where the most frequently occurring patterns are selected by considering the occurrence of patterns in the training images. Hence, the goal is to learn the subset of discriminative patterns from the training images dataset (patches) and retain the patterns based on their distribution for the classification task. Therefore, also for a complex image, the patterns with a high occurrence frequency will be selected. The dictionary of the most frequent patterns can be computed by summing together all the RLBP histograms ($H_i \in \mathbb{N}^{2^P}$) and considering the complete possible set of RLBP values from the training images ($I_1, I_2, \ldots, I_T$); then, the resulting histogram ($H = \sum_1^T H_i$) is sorted in descending order to select only the patterns corresponding to the first $M$ bins as shown in Fig. 8. Where the selected $M$ patterns depend on the threshold parameter ($\theta$) and the training data. $M$ can be computed by Eq. 6.

$$M = \arg \min_m \left( \frac{\sum_{i=1}^{M-1} H_{\text{sorted}}[i]}{\sum_{i=1}^{2^P} H_{\text{sorted}}[i]} > \theta \right) \tag{6}$$

Finally, the new feature histograms on the selected patterns for channels (H and V) are concatenated to form the final feature vector, and then an image classification based on the DRLBP extracted patterns is performed.

## Classification

In this section, we are interested in learning methods in a description space as defined in the previous section. The goal of this step is to build a decision function from learning data projected in this description space for the binary patches classification as tumor or not-tumor. There are two main trends in learning methods: the single classifier approach and the approach with a combination of classifiers aggregating the responses of several classifiers Breiman96.

The first approach consists of developing a unique decision function directly from the training data by estimating a particular distribution of classes (these approaches are called generative) or by drawing a decision boundary separating the classes (these approaches are called discriminative). In this work, we first apply the current approaches of the literature in the context of learning a single classifier like support vector machines (SVM) (Vapnik et al. 1996), k-nearest neighbor (k-NN) (Mucherino et al. 2009). In a second step, we develop approaches by combining classifiers. We particularly interest in homogeneous ensemble methods based on decision trees (random forest (RF) (Breiman 2001)).

The current contribution of deep neural networks allowed great performance on computational histopathology (Srinidhi et al. 2019; Dimitriou et al. 2019), which led us think that it is worth to experiment and test the convolutional neural networks (CNNs) for the binary patches classification as tumor or not-tumor. Deep neural network, unlike other methods, builds its own analysis features. Thus, it can identify high-level features from data incrementally, and thereby completely eliminating the manual feature extraction phase. However, to achieve satisfactory performances these models rely on huge datasets to train a model, but in some contexts, it is very difficult, and sometimes impossible, to have a large dataset to train a model, as is the case of our study with only 12 WSI histological image datasets with 1800 labeled patches.

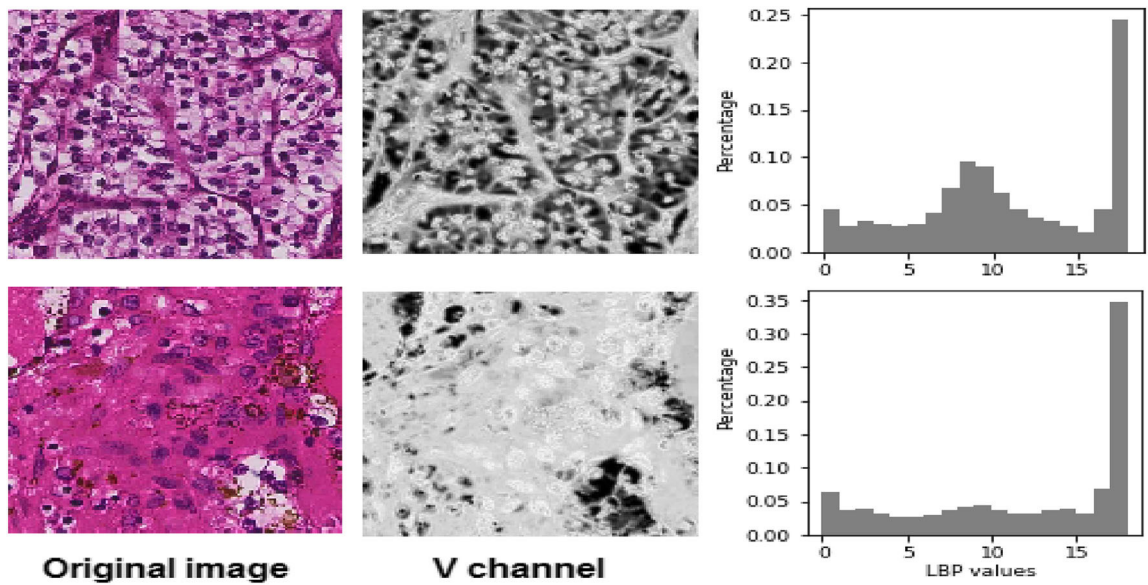

**Fig. 7** Histograms of LBP patterns with ($P = 16$) neighbors for each pixel on the image. First row: tumor patch. Second row: not-tumor patch

In this case, transfer learning allows to speed up network training and helps prevent over-fitting. Indeed, when the collection of input images is small, it is strongly advised not to train the deep neural network starting from scratch (with random initialization): the number of parameters to be learned being much greater than the number of images, the risk of over-fitting is very important.

With transfer learning, we can exploit the pre-trained neural network in two ways (deep feature extraction and partial fine-tuning), depending on the size of the input dataset.

In our work, since the dataset contains only 12 WSI histological image datasets with 1800 labeled patches, which represents a small number of images, the Deep feature extraction may be a good solution by using the features of the pre-trained network to represent them knowing that the features of the lower layers are simple and generic (so they can be found in two very different images), while the features of the upper layers are complex and problem-specific. Thus, the strategy of fixing the lower layers and training the classifier and the upper layers is a good compromise. On the other hand, the partial fine-tuning approach, where the last fully connected layer is replaced again by the new randomly initialized classifier, and the parameters of some layers of the pre-trained network are fixed. This approach can meet what we are looking for in the applied whole slide images of renal cell carcinoma.

To explore these two solutions, we propose the use of transfer learning of the pre-trained models on the giant ImageNet dataset (Deng et al. 2009), like ResNet (He et al. 2016) and VGG (Simonyan and Zisserman 2015), since the training from scratch of a deep convolutional neural network is a very challenging task with very small size image dataset by causing over-fitting. Indeed, the pre-trained VGG and ResNet on the ImageNet dataset learn thousands or millions of parameters to efficiently identify images. Therefore, we experiment two transfer learning strategies: deep feature extraction and partial fine-tuning with pre-trained VGG and ResNet model.

## Results

### Image dataset

To evaluate the method outcomes, we used a collected dataset of 12 hematoxylin and eosin (H & E)-stained whole slide images of kidney tumors (RCC) gathered from the University Hospital Center of Nice in France, which were collected from different patients with a diagnosis of RCC (renal cell carcinoma) from each of the two main malignant tumor types: clear cell renal cell carcinoma (ccRCC) and papillary carcinoma (pRCC). The WSIs were digitized at $400\times$ magnification by a Leica SCN400 slide scanner then automatically compressed in "scn" format (whole slide

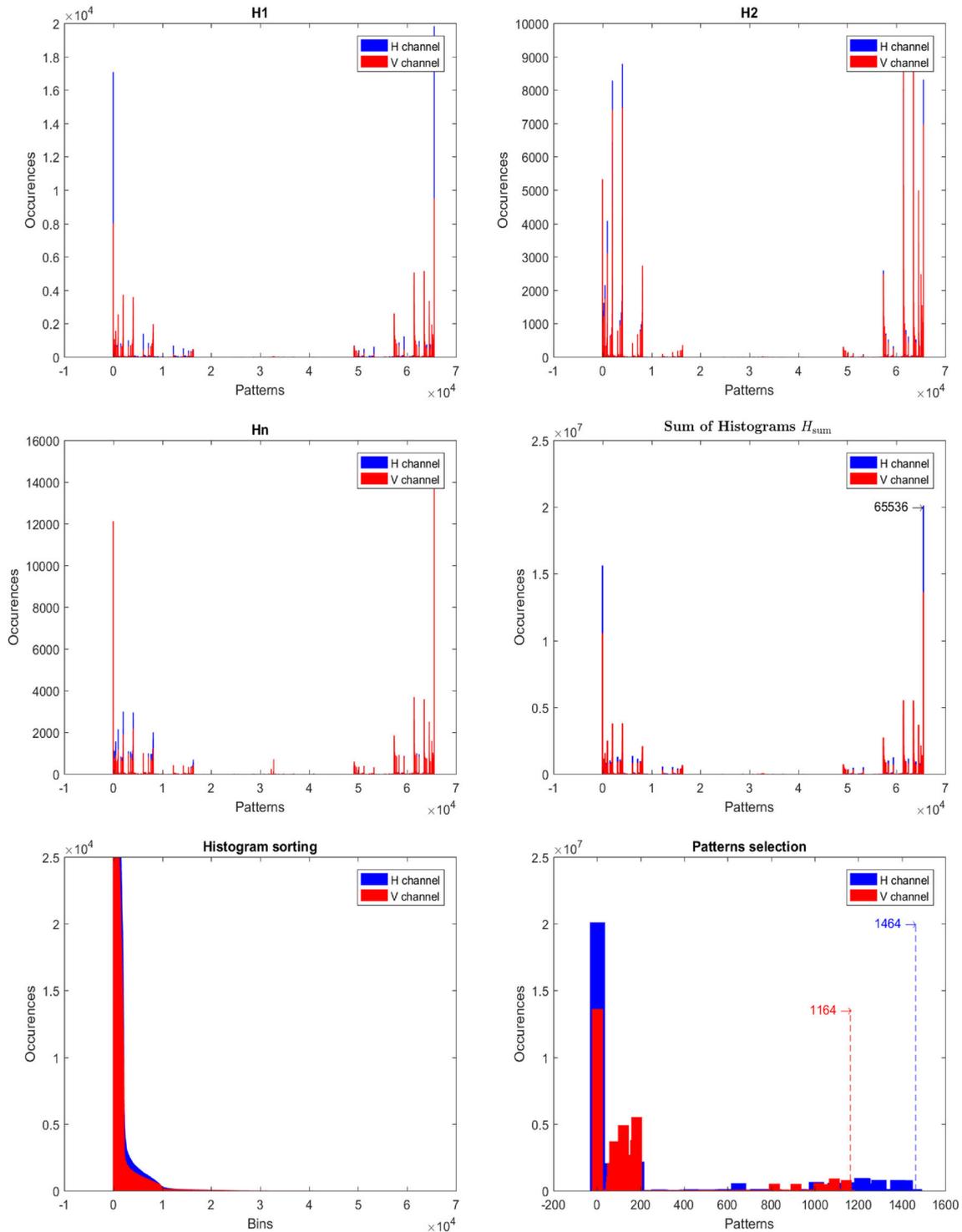

**Fig. 8** Feature selection. (From top left to bottom right) RLBP histograms ($H_1, H_2, \ldots H_n$) for training images where $P = 16$ and the initial total number of patterns for each channel (H and V) is 65536; sum of histograms $H_{sum}$; summed histogram bins sorted in decreasing order for selection of the most frequent patterns (the shown figure is zoomed); histogram of selected patterns. A total of 1464 and 1164 patterns are selected (which is considerably low compared to the initial descriptor dimensionality) from the H and V images respectively corresponding to the threshold parameter ($\theta = 0.90$)

imaging) for observing tissue at multiple magnification levels.

The first level allows the observation of the whole tissue at low magnification to distinguish the different tissue types (healthy, tumor, fat, necrosis, blood, etc.) as can be seen from Fig. 3 (leftmost), while the highest magnification level (400×) enables a cellular scale view of the tissue as shown in Fig. 3 (rightmost). The provided slide images have a resolution of 0.25 m with a total of about 100.000 pixels per axis at 400× magnification, Figs. 1 and 3 show thumbnail colored images of this dataset, and hence, the entire WSI is divided into color patches (RGB) of (600 × 600) size pixels at the highest resolution for analysis.

The ground truth images are generated to evaluate the performance of the proposed method. Firstly, the slides have been manually annotated by an expert as a region containing an ROI (regions of interest), since the entire tumor area is not of interest, like necrosis. Consequently, each WSI has an associated ground truth image where the ROIs are roughly surrounded at low magnification level (see Fig. 1). We randomly partitioned 8 of these slides (two-thirds of the dataset) for training and the remaining 4 slides (one-third of the dataset) for testing. To counteract computational difficulties, a total of 1800 patches, of size 600×600, were extracted from the different scans (slides) at 400× magnification level. Thereafter, for the training set, pathologists manually annotated 1.200 patches from the 8 slides, about 150 patches per slide. These patches have the same size (600 × 600) and were labeled as either "tumor" or "not-tumor," for our experiments, we grouped the ccRCC and the pRCC patches into a single class named "Tumor". Therefore, the patch-level label includes two classes: (0) not-tumor and (1) tumor. For the test set, the pathologists annotated 600 patches of size 600×600 from the 4 slides, about 150 images per slide, ensuring that patches from the same slide are not present for training and testing. Moreover, the distribution of classes is balanced in the training and test set to avoid biasing the model towards a particular class that has the most samples.

## Experimental setups

Our approach based on patch classification with a sliding window to identify the ROIs on a given whole slide image, where a patch extraction step using the nuclei ratio was introduced as a preprocessing step. Then, to classify the selected patches into two classes (tumor or not-tumor) using three classifiers (k-NN, SVM and RF), we calculated the DRLBP features of all patches, converted to H channel (hematoxylin) and V channel (violet), from the circularly symmetric neighborhoods by varying the parameters $P$ (number of neighbors) and $R$ (radius) with values of $(P = 8, 12, 16)$ and $(R = 1, 2, 3)$. The classification accuracy saturates after multiple tests with a significant reduction in feature dimension with the threshold value $\theta$ between 0.85 and 0.90. For this reason, we opt for the typical (P, R) values of (16, 3) and $\theta = 0.90$ to capture discriminative information, since larger $(P, R)$ values increase considerably the dimensionality and computational complexity without significant performance improvements. In the same way, the accuracy further decreases when increasing the values of $\theta > 0.95$. That is explained by the selection of non-discriminative patterns with a high threshold value (Mehta and Egiazarian 2016). As a result, the implemented method improves both the classification accuracy and significantly reduces the feature dimensions with a range of selected patterns within 2 to 5% of the overall pattern number. In our case study, we select around 4% (1164 and 1464 patterns from a total set of 65536 patterns) representing 1.77% and 2.23% for the V and H channels respectively; this is illustrated in Fig. 8. For evaluation, we use different classifiers: k-NN with a typical value of $k = 5$, SVM based on the radial basis function (RBF) kernel, and random forest (RF) that requires as inputs the number of trees forming the forest, during our experiments and after multiple tests we opt for the value of 1000 trees. Furthermore, we implemented a grayscale patch classification scheme to compare results, and Table 1 summarizes the classification performance.

The most common metrics for image classification as Accuracy, Precision, IoU and Dice Index are applied to evaluate the proposed approach, these metrics are given as follows:

- **Accuracy** Distance between a measurement and reality.
- **Precision** is the ability to be able to repeat measurements with the same result. It measures the homogeneity of the data, that is, the ability to be able to predict the behavior of the algorithm's error. Indeed, if the reliability is 100%, then the error in precision of the algorithm is the same for the whole image.
- **Jaccard Similarity Index (IoU)** It allows to calculate the similarity between two shapes (an image $U$ and ground truth $V$) by determining the area they have in common compared to their total area. It is often used to rank algorithms in different medical imaging competitions.

$$IoU(U, V) = 100 \times \frac{|U \cap V|}{|U \cup V|} \qquad (7)$$

- **Dice Similarity Index (Dice)** Similar to the Jaccard index defined in Eq. 7, it is also an index of similarity between two objects and it is used in many scientific contributions. The main difference is that, unlike the Jaccard index, it does not satisfy the triangular

**Table 1** Classification results for histopathology test dataset

| Methods | | Accuracy | Precision | Dice | IoU |
|---|---|---|---|---|---|
| k-NN | RLBP H&V | 98.50 | 98.21 | 98.19 | 96.45 |
| | LBP H&V | 96.50 | 96.68 | 96.61 | 93.44 |
| | RLBP Gray | 97.83 | 97.89 | 97.87 | 95.83 |
| | LBP Gray | 96.33 | 96.46 | 96.43 | 93.10 |
| SVM | RLBP H&V | **99.17** | **99.17** | **99.17** | **98.36** |
| | LBP H&V | 98.33 | 98.34 | 98.34 | 96.74 |
| | RLBP Gray | 98.83 | 98.84 | 98.84 | 97.71 |
| | LBP Gray | 98.17 | 98.17 | 98.17 | 96.41 |
| RF | RLBP H&V | **99.00** | **99.01** | **99.01** | **98.03** |
| | LBP H&V | 97.17 | 97.21 | 97.21 | 94.57 |
| | RLBP Gray | 98.83 | 98.84 | 98.84 | 97.71 |
| ResNet-50 | Fine-tuning | - | 98.50 | 98.14 | - |
| VGG16 | Deep Features | - | 95.50 | 95.00 | - |

Bold font indicates the highest performance inequality stating that $|U + V| \leq |U| + |V|$. Indeed, its denominator is greater than or equal to the sum of the distances $|U + V|$, which can cause a risk of overlapping areas. This implies that the Dice index will overestimate the true similarity value. It is therefore interesting for comparing results with each other but does not allow a similarity result to be validated.

$$Dice(U, V) = 100 \times \frac{2 \times |U \cap V|}{|U| + |V|} \qquad (8)$$

For CNN's experimentations, the first model used for detection of ROIs in our histological WSI was a partial fine-tuning ResNet-50 pre-trained on the ImageNet dataset, by freezing the first blocks until "*res5a_branch2a*" (i.e., block "5," sub-block "a") and training only the last few layers, in order to transfer the primitive low-level features learned on ImageNet by the first blocks layers, and then learned only the high-level features specific to our dataset images. Finally, adding our fully connected layer with 1024 neurons, the dropout layer (with a rate of 0.5) then a single dense layer with sigmoid activation function as our classifier. The dataset was split into training, validation, and testing, where patches were resized from 600×600 to 224×224 for inputs. Additionally, we performed data augmentation (rotation 90°, shifting, horizontal and vertical flipping) on the training images, this helps the model generalize better. We stopped the training after 16 epochs when the validation accuracy failed to improve. We obtained 98.50% (see Table 1) precision on the test set.

Regarding the deep feature method based on VGG-16 architecture. We instantiate only the convolutional part; everything up to the fully connected layers of the VGG-16 model pre-trained on the ImageNet dataset. By next, we run this model on our histological images (patches) with a patch size of 224×224 pixels, the output layer (bottleneck features) from the architecture will be an array with a shape of 7×7×512 and we get features extracted arrays. Finally, the output is reshaped and trained with classic neural networks for binary classification (tumor, not-tumor). This model achieved a test precision of 95.50% (Table 1 presents results). All the models were implemented and trained on Colab (Google Colaboratory) GPU environment using TensorFlow with a Keras library.

## Discussion

In this paper, we present an automatic system that can distinguish tumorous from normal tissue using histopathology images. Experimental evaluation of our proposed approach relative to expert manual annotations demonstrates its efficacy in both visual and quantitative measurements. Fig. 9 illustrates results for ROI identification samples, where the first column shows the input images, the second one is ground truth images and the third indicates the detected regions of interest (foreground). The proposed approach yields excellent performance for the identification of ROIs (regions of interest) in most cases even if they are spread across the whole slide image (WSI) since the background and various non-tumor patches (debris, blood, fat tissue, etc.) are completely removed from the image in the pre-processing step and the small areas are removed using morphological operations. In addition, secondary regions of eventual unhealthy tissue are detected as shown in the second row of Fig. 9. Furthermore, it should be noted that better performance is achieved when using a 400× magnification to distinguish non-tumor from tumor regions which typically have varying and complex morphological patterns at the microscopic level (nucleus and vessel features) (Srinidhi et al. 2019; Tabibu et al. 2019).Indeed our preprocessing step focuses on the detection of thousands of nucleus cells using pixel-wise and sliding window methods for patch extraction. We can potentially achieve better results by using resolutions conjointly; however, analyzing slides at different magnification levels using various methods increases both complexity and processing time. Nevertheless, some difficulties encountered are mainly due to the blurred areas that can be found in some images as shown in Fig. 12.

**Fig. 9** ROIs identification results (from left to right): input images, ground truth image, detected regions of interest. (The aspect ratio has not been respected when displaying some images)

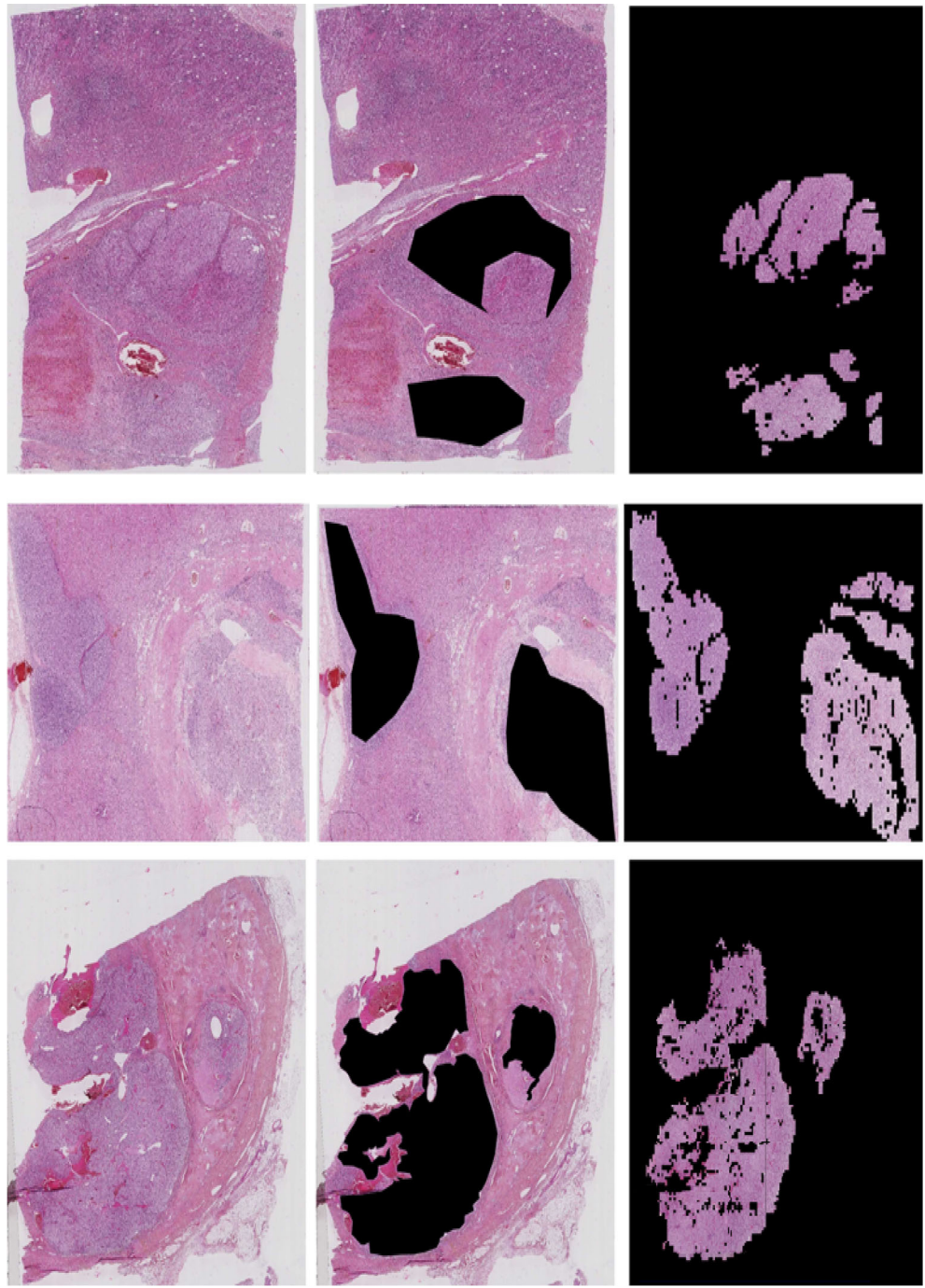

For a quantitative comparison, we performed image classification on the testing set with multiple classifiers (k-NN, SVM and RF) using the standard LBP (Local Binary Pattern) and the proposed DRLBP features by concatenating various configuration of the extracted features (H cannel, V channel and gray level images) with the same parameter ($r = 3$ and $P = 16$) as shown in Table 1. It can be observed that the proposed DRLBP descriptor achieves the highest performance with precision of 99.17% using SVM. In addition, the results in Table 1 show that the RF and

**Fig. 10** Segmentation results using k-means for different *k* values (Zubiolo 2015)

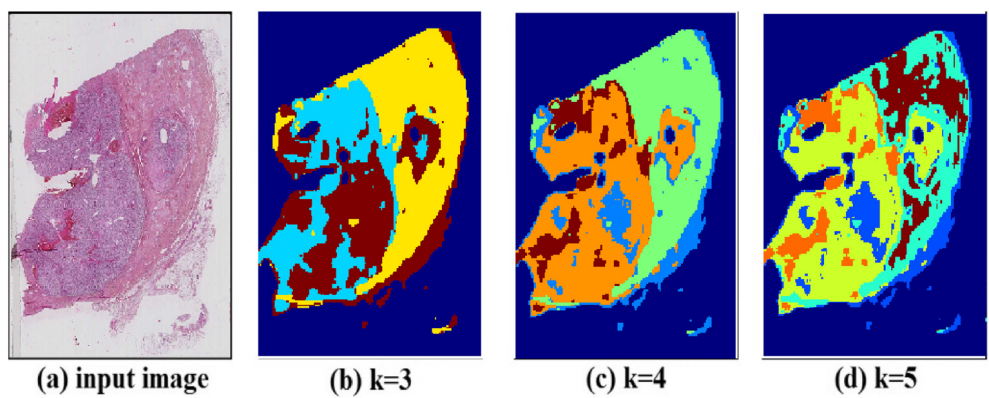

(a) input image     (b) k=3     (c) k=4     (d) k=5

k-NN classifiers generally get better results comparing to the LBP descriptor, which proves that the modifications on rotation changes provide discriminatory patterns. Further, it is also interesting to observe that the DRLBP using H channel and V channel achieves significantly higher accuracy than the DRLBP with the same parameters on grayscale image, which indicates that the operator captures region-based information is more effective and the concatenated features performed well in representing histopathological texture patterns. Nevertheless, the SVM-DRLBP classification is quite fast, taking 77 ms to process one patch, the k-NN–DRLBP classification requires 87 ms and the RF–DRLBP classification requires 1.55 s/patch. Considering the experiments have all been performed on a machine with the following characteristics (CPU: INTEL Core(TM) i7-3632 2.20 GHz, RAM 12GB).

Table 1 compares the performance of the proposed approach with the CNNs transfer learning namely ResNet-50 and VGG-16. The results were very promising with the precision of classifying tumors and not-tumor patches of around 98.50% for ResNet-50 and 95.50% for VGG-16. These models demonstrated that transfer learning was also efficient and achieved a good performance when dealing with a small image dataset. Nevertheless, the fine-tuned ResNet-50 architecture performs better than deep features extraction VGG-16, since the ResNet-50 last layers were trained on our histological image dataset. These results revealed that the proposed DRLBP approach can classify patches and identify the ROIs with high accuracy, where complex and pre-trained CNNs models with millions of parameters have comparable performance on a small histopathological image dataset.

Additionally, a qualitative comparison was conducted with Zubiolo's (2015) method cited in the related work, where the same images have been employed to identify ROIs and the segmentation results are shown in Fig. 10, it can be seen that the regions of interest (surrounded in black in Fig. 11 b) are identified by the proposed scheme for reasonably low *k* value. However, a high *k* value (see Fig. 10 d) leads to over-fitting and therefore requires selecting several classes to segment the region of interest (i.e., *k* = 3 or *k* = 4).

This approach, although giving interesting results for some images (Zubiolo 2015), presents some limitations and requires improvement for systematic application. Indeed, using an unsupervised learning approach to automate the detection of ROI is a challenging task due to the intra-image and inter-image variability where the k-means algorithm is repeated *m* = 3 times to have different partitions according to the centroids initialization; then, a majority vote is

**Fig. 11** Segmentation results using *k-means* (Zubiolo 2015) and our proposed approach. Regions of interest indicated in black on the ground truth image

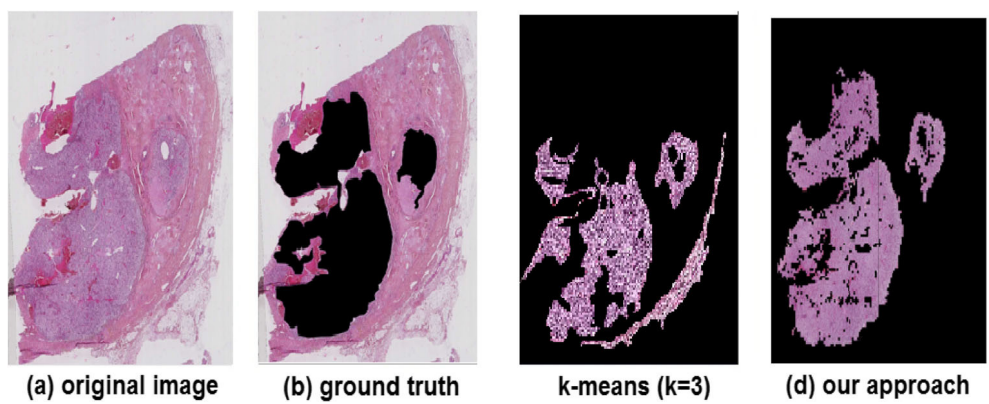

(a) original image     (b) ground truth     k-means (k=3)     (d) our approach

**Fig. 12** Example of misfiled image where blurry bands appear in the whole slide image

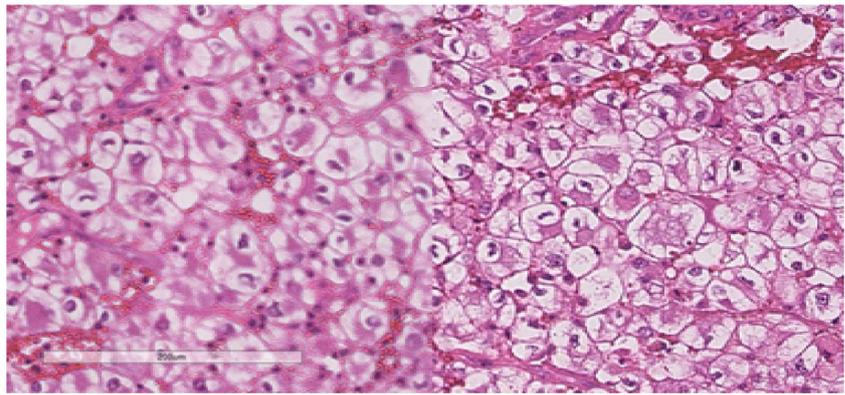

required for the final pixel partition. Moreover, the used features (entropy, variance, and median) capture only pixel information, unlike our proposed approach (see Fig. 11) where the region-based information is captured with image patches rather than the entire WSI for performance enhancement without additional computational costs.

The main emphasis of this study is to exploit and integrate visually meaningful features (texture and color) to develop an automated system based on general and domain-specific features leading to more accurate results in RCC segmentation. The idea behind the proposed method is that the measurement of texture characteristics in eosin and hematoxylin-stained slides can greatly improve the DRLBP recognition accuracy when dealing with wide pattern variability in histopathological whole slide images, which is robust to image rotation, grayscale changing (linear function) and insensitive to noise and histogram equalization. DRLBP combined with color transformation (H and V channels) has been applied to retain the structural information, exploit the magnitude values in a local neighborhood for more discriminative power and reveal the immense texture variability at the microscopic high magnifications level.

Despite the above-mentioned strengths, several possible pitfalls are noteworthy for discussion. As shown in Fig. 12, during the slide scanning process, some images contain blurred bands due to incorrect automatic sensor focus. In such a case, even for the human expert, it's difficult to recognize the different image regions. In practice, the pathologists simply ignore these regions since the information that is contained is too weak and there are sufficiently vast areas to explore owing to the large image size. Therefore, the lab technician in charge of scanning spends a lot of time manually checking every slide after scanning, and re-scans corrupted slides, making it a tedious and expensive procedure.

## Conclusion

In this paper, an automatic system for tumor region identification in large-scale histopathology images at the microscopic level has been proposed to assist pathologists to diagnose in kidney cancer and reduce significantly their workloads. The proposed workflow exploits the texture and color characteristics of the whole slide images by using the dominant rotated local binary pattern (DRLBP) descriptor and reducing the dimensionality of the features. We noticed that our simple and robust approach provides excellent results for tumor region identification in histological images, which is mainly due to the DRLBP features extraction from different color channels (H and V) providing rich texture information and the features selection to select the majority of texture patterns that can well describe the texture characteristics of histological images to provide a final feature vector through concatenation. Indeed, the DRLBP highlights the non-homogeneous regions of the histological images and preserves more texture information by capturing discriminative information in the features histograms by discarding the non-discriminatory patterns. Furthermore, our experiment reveals possible unhealthy tissues during the classification task which is potentially due to the formation of secondary tumor regions around the primary site (regional metastases). A comparative analysis was conducted using CNNs transfer learning models and the results reveal comparable performances, this suggests that our approach has great potential and the possibility for application in other related problems such as breast cancers where the size of the datasets is relatively small. Based on these promising results, in our future research, we intend to concentrate on tumor morphological features extracted from the final image and the classification of renal cell carcinoma (RCC) subtypes to build a prognostic model and predict survival outcomes.

**Acknowledgements** The authors would like to thank the Directorate-General of Scientific Research and Technological Development (Direction Générale de la Recherche Scientifique et du Développement Technologique, DGRSDT, URL: www.dgrsdt.dz, Algeria) for the financial assistance towards this research.

## Declarations

**Ethics approval** For this type of study, formal consent is not required.

**Consent to participate** This article does not contain any studies with human participants or animals performed by any of the authors.

**Informed consent** This article does not contain patient data.

**Conflict of interest** The authors declare no competing interests.

## Affiliations

**Mohammed Lamine Benomar[1,2]**  **· Nesma Settouti[1] · Eric Debreuve[3] · Xavier Descombes[3] · Damien Ambrosetti[4]**

Nesma Settouti
nesma.settouti@univ-tlemcen.dz

Eric Debreuve
eric.debreuve@inria.fr

Xavier Descombes
xavier.descombes@inria.fr

Damien Ambrosetti
ambrosetti.d@chu-nice.fr

[1] Biomedical Engineering Laboratory, University of Tlemcen, Tlemcen, Algeria

[2] University of Ain-Temouchent, Belhadj Bouchaib, Ain-Temouchent, Algeria

[3] Université Côte d'Azur, CNRS, Inria, I3S, Nice, France

[4] Laboratoire Central d'Anatomo Pathologie, CHU Nice. Hôpital Pasteur, Nice, France